\documentclass[11pt]{article}
\pdfoutput=1
\usepackage{jheppub}

\DeclareFontFamily{U}{wncy}{}
\DeclareFontShape{U}{wncy}{m}{n}{<->wncyr10}{}
\DeclareSymbolFont{mcy}{U}{wncy}{m}{n}
\DeclareMathSymbol{\Sha}{\mathord}{mcy}{"58} 
\def\unit{{\hbox{\kern+.5mm 1\kern-1mm l}}} 

\usepackage{graphicx}
\usepackage{subcaption}
\usepackage[english]{babel}
\usepackage{braket}
\usepackage{amsmath}
\usepackage{amssymb}
\usepackage{bbold}
\usepackage{bbm}
\newcommand{\df}{\textnormal{d}}
\newcommand{\dbraket}[1]{\braket{\braket{#1}}}
\newcommand{\op}[1]{\mathcal{#1}}
\title{\textbf{Classical Open String Amplitudes from Boundary String Field Theory}}

\abstract{We calculate the corrections to the open string BRST charge in boundary string field theory and show that the expansion coefficients agree with the perturbative S-matrix to all orders. In mathematical terms, boundary string field theory describes the minimal model of open string field theory.}

\author{Christoph Chiaffrino,}
\author{Ivo Sachs}
\affiliation{Arnold Sommerfeld Center for Theoretical Physics,\\ Ludwig-Maximilians-Universit\"at,\\
Theresienstr. 37, D-80333 M\"unchen, Germany}
\emailAdd{Christoph.Chiaffrino@physik.uni-muenchen.de}
\emailAdd{Ivo.Sachs@physik.uni-muenchen.de}

\begin{document}
\maketitle
\section{Introduction and Summary}\label{IS}

The standard formulations of string theory rely on a choice of background, more precisely, a choice of conformal field theory. Different backgrounds can be realized, for example, by a curved target space manifold. In this case strings move in a curved spacetime. For open strings there is an additional freedom in the choice of a boundary condition, which one usually thinks of as certain brane configurations. Different choices lead to inequivalent theories in general; one says that these theories are background dependent.

It has been known for some time \cite{Sen:1990hh,Sen:2017szq} that it is possible to access nearby backgrounds from a specific one. On general grounds, the problem of background deformation covariant string field theory is well posed when the Hilbert space is preserved by the perturbation \cite{Munster:2012gy}. This is the case, in particular, for open string field theory on a fixed closed background (i.e. bulk CFT) and boundary perturbations generated by operators obtained from the bulk CFT. We then expect that a background independent formulation of string theory should exist.

In \cite{Witten:1992qy} Witten proposed a formulation for a background independent classical bosonic open string theory. As reviewed in section \ref{BIS}, this theory is formally defined as a Batalin-Vilkovisky (BV) action $S$ on the space $\mathcal{F}$ of two dimensional boundary conformal field theories (BCFT) on the disk. Classical backgrounds correspond to extrema of $S$.  This theory is commonly known as boundary string field theory (BSFT).

In \cite{Witten:1992qy} the space $\mathcal{F}$ is parametrized by the set of boundary interactions,  $ \oint \mathcal{V}$ of local operators, $\mathcal{V}$.\footnote{Non-local boundary interaction were argued in \cite{Baumgartl:2004iy} to correspond to a shift in the closed string background.}  In concrete calculations one is generically forced to treat the $\mathcal{V}$ as a perturbation around a conformal background. In this setup Shatashvili showed \cite{Shatashvili:1993ps} that contact terms arising in the perturbative expansion spoil the naive proof of gauge invariance. He proposed a modification in terms of deformed Virasoro generators (see also \cite{Sen:1990hh}), for which he proved gauge invariance to first order in perturbation theory. These contact terms give rise to nonlinear equations of motion. 

In section \ref{s:onshell}, we will argue that for on-shell perturbations these non-linearities agree precisely with the tree-level perturbative string $S$-matrix to all order in perturbation theory, that is 
%
\begin{equation}\label{Action}
S(\phi) = \sum_{n \ge 3} \frac{1}{n} \op{V}^{(n)}_S(\phi, ...,\phi)\,,
\end{equation}
where $\mathcal{V}^{(n)}_S(\phi,...,\phi)$ denotes the symmetrized $n$-point tree-level scattering matrix.

In the mathematical literature (e.g. \cite{Kajiura:2003ax}) classical $S$-matrices are synonymous to the minimal model of the underlying $A_\infty$ algebra of a BV-action.\footnote{In mathematical literature this construction is known as the homological perturbation lemma.}  In the case at hand this  $A_\infty$ algebra is that of open bosonic string field theory (OSFT) \cite{Witten:1985cc}. An equivalent interpretation of our result  (\ref{Action}) is then that BSFT, restricted to physical states,  reproduces the minimal model \cite{Kajiura:2003ax} of open bosonic string field theory. 

This is different from the usual way of computing homological minimal models, since the latter is constructed in terms of Feynman diagrams derived from the vertices of the classical OSFT action.
Given that BSFT parametrizes the minimal model of OSFT one may ask what the minimal model of BSFT is. In section \ref{s:offshell} we will argue on general grounds that there are no vertices in $S$ that connect on-shell states to a single off-shell state. Consequently, the only contributions to tree-level S-matrices are given by contact vertices. Thus, the minimal model of BSFT is again that of OSFT. This also shows that around a conformal background, BSFT and OSFT are perturbatively equivalent since their scattering amplitudes agree.

\section{Background Independence, Perturbative Expansions and the Minimal Model in String Field Theory}\label{BIS}

It is assumed that all string field theories possess the structure of a Batalin-Vilkovisky (BV) theory. It is given by an action $S$ defined on a graded field space $\mathcal{F}$. Ignoring quantum effects the action satisfies the classical master equation
\begin{equation}\label{meq}
\{S,S\} = 0.
\end{equation}
The odd Poisson bracket $\{\cdot, \cdot\}$ is defined through an odd symplectic structure $\omega$ in the usual sense. The action $S$ defines a Hamiltonian vector field
\begin{equation}\label{VQ}
V = \{S,\cdot\},
\end{equation}
which generates gauge transformations $\delta f$ of functionals $f$ on $\mathcal{F}$ through
\begin{equation}
\delta f = V(f).
\end{equation}
In particular, the action itself is gauge invariant because of (\ref{meq}). Equation (\ref{VQ}) can also be restated in the form
\begin{equation}
\delta S = i_V\omega.
\end{equation}
Therefore, zeros of $V$ are critical points of $S$ ($\omega$ is assumed to be non-degenerate). The BV data is usually denoted by the triple $(\mathcal{F}, \omega, V)$.

$V^2 = 0$ implies that the vector field $V$ defines a cohomology on the functionals on $\mathcal{F}$. This cohomology represents gauge invariant functionals modulo gauge equivalences. One can transfer this cohomological structure to the tangent space $T_p\mathcal{F}$ by a Taylor expansion
\begin{equation}\label{VQexp}
V_p = V_p^{(0)} + V_p^{(1)} + ...\, .
\end{equation}
around a point $p$ in field space. The $V_p^{(n)}$ define multilinear maps
\begin{equation}
V_p^{(n)}: (T_p \mathcal{F})^{\otimes n} \rightarrow T_p \mathcal{F}.
\end{equation}
If $p$ is a critical point, then $V_p^{(0)} = 0$.  The maps $V_p^{(n)}$ together with $\omega_p$ then define the vertices of order $n+1$ of the string field theory action $S$ expanded around the point $p$. 
These vertices define a cyclic $A_\infty$-algebra on $T_p \mathcal{F}$ which makes the perturbative gauge invariance of the action $S$ manifest. 


The linear term
\begin{equation}
V_p^{(1)} : T_p \mathcal{F} \rightarrow T_p \mathcal{F}
\end{equation}
in (\ref{VQexp}) encodes the free equations of motion as well as the linear part of the gauge transformations of the theory expanded around the point $p$. From $V_p^2 = 0$ it follows also that $V_p^{(1)} \circ V_p^{(1)} = 0$. With this we can also define the cohomology $H = H(T_p \mathcal{F},V_p^{(1)})$ given by the equivalence classes of on-shell fields modulo linear gauge transformations. They are the asymptotic states of the theory.

One can transfer the $A_\infty$ structure on $T_p\mathcal{F}$ to a {\it minimal} $A_\infty$ structure on $H$, essentially by evaluating the Feynman diagrams with the vertices  $V_p^{(n)}$. We denote the multilinear maps of the minimal algebra by ${V}_{\text{min}}^{(n)}: H^{\otimes n} \rightarrow H$, where now, in addition,  $V^{(1)}_\text{min} = 0$. At ghost number zero they represent the tree-level S-matrices, in the sense that the $n$-point scattering of states $v_1,...,v_n$ is equal to
\begin{equation}
\mathcal{V}^{(n)}(v_1,...,v_n) = \frac{1}{n}\omega_p(v_1,V_\text{min}^{(n-1)}(v_2,...,v_n)).
\end{equation}

We can summarize the above steps in the following diagram
\begin{equation}
(\mathcal{F}, \omega, V) \stackrel{\text{crit. point}}{\longrightarrow} (T_p\mathcal{F},\omega_p,\{V_p^{(n)}\}) \stackrel{\text{min. model}}{\longrightarrow} (H(T_p\mathcal{F},V_p^{(1)}), \omega_p,\{V_{\text{min}}^{(n)}\}).
\end{equation}
Historically, this deductive construction developed backwards, and also only partially so. The original string theory is a theory of S-matrices only. It is defined on-shell (i.e. on $H(T_p\mathcal{F},V_p^{(1)})$) and as such the $\{V_{\text{min}}^{(n)}\}$. The versions of string field theory developed afterwards are from this viewpoint a triple $(T_p\mathcal{F},\omega_p,\{V_p^{(n)}\})$, from which one can deduce string theory amplitudes using Feynman diagrams. Those string field theories still rely on an a priori choice of a background (a critical point $p$).

Boundary String Field Theory (BSFT) was developed in \citep{Witten:1992qy} to give a first step towards describing $(\mathcal{F}, \omega, V)$. It is a string field theory that is formally independent of the open string background. In practice this means that while we have to choose a CFT to describe the propagation of strings in the bulk, its definition does not rely on a specific worldsheet boundary condition (conformal or not).

Despite its background independent definition, calculations in BSFT have so far only been done using conformal perturbation theory. In our language this means that one directly calculates the $V_p^{(n)}$ without having a closed expression for $V$. Using this perturbative framework it was noticed in \citep{Shatashvili:1993ps} that certain assumptions made in the original paper cannot hold. One of them was the independence of $V_p$ from boundary conditions, although well motivated since $V_p$ was defined through the fixed bulk CFT, it gets spoiled because of contact terms. In hindsight this observation proved to be advantageous for BSFT as a field theory of open strings, since otherwise the theory would be linear, as was shown in \citep{Shatashvili:1993kk}. The drawback is that we face the usual arbitrariness appearing when one has to introduce counter-terms to cancel infinities. This makes the definition of the $V_p^{(n)}$ ambiguous. In \citep{Shatashvili:1993ps} a choice was made to fix this ambiguity and with this the calculation was done to compute $V_p^{(2)}$.

In this work we want to motivate another way to deal with contact terms, which is well defined for on-shell perturbations. It has the advantage that it allows us to determine the $V_p^{(n)}$ to all orders of $n$ in one computation.  We will show that with this prescription, BSFT reproduces string theory S-matrices and it does this in a very special way: We will see that, while restricting ourselves to on-shell perurbations, the vertices of the theory are the string theory S-matrices. This means that the vertices of BSFT reproduce the minimal model of open string field theory. Moreover we will argue why these vertices are also the minimal model of BSFT, i.e.
\begin{equation}
V_p^{(n)}(v_1,...,v_n) = V^{(n)}_{\text{min}}(v_1,...,v_n).
\end{equation}

\section{Construction of the BSFT Action}

\label{s:onshell}

As explained in the previous section, boundary string field theory is by definition a BV action. As such its definition depends on the following ingredients:
\begin{enumerate}
\item A symplectic form $\omega$ of degree $-1$. In particular this means $\omega$ should be closed. \label{prop1}
\item A degree $1$ vector field $V$ which squares to zero. Such a vector field is called cohomological. \label{prop2}
\item The vector field $V$ generates a symmetry of $\omega$. \label{prop3}
\end{enumerate}
Property \ref{prop1} and \ref{prop3} imply that $i_V\omega$ is closed:
\begin{equation}
0 = L_V\omega = (d \circ i_{V} - i_{V} \circ d)\omega = d(i_V\omega).
\end{equation}
This allows us to define locally an action $S$ via $\df S = i_{V}\omega$. Property \ref{prop2} then implies that $S$ satisfies the classical master equation $\{S,S\} = 0$ which is equivalent to $S$ being gauge invariant under the transformations generated by $V$.

In the context of BSFT \cite{Witten:1992qy} these objects were identified as follows: First of all the space of fields ${\cal{F}}$ is  the space of all two dimensional sigma models defined on the disc with fixed bulk CFT action $I_0$. Points in ${\cal{F}}$ are represented by boundary operators $\mathcal{O}$ defined through the bulk. The complete world sheet action is then
\begin{equation}
I = I_0 + \oint_{S^1} \df s \, b_{-1}\mathcal{O}(s)\,, 
\end{equation}
where $b_{-1}$ is defined in terms of the closed string antighost field (we come to its proper definition later). We parametrize our boundary perturbation by spacetime fields, so that $\op{O} = \phi^i \op{O}_i$. When we do conformal perturbation theory we think of the $\mathcal{O}(s)$ to be small and in this sense the $\mathcal{O}_i$ are a basis of the tangent space. Like in open string field theory the statistics of the fields are such that $\op{O}$ is fermionic, this will also be necessary to have a bosonic worldsheet action $I$.

The symplectic form at a point $\mathcal{O}$ is the expectation value of two tangent vectors taken with respect to $I$:
\begin{equation}\label{symplform}
\omega(\delta_1\op{O}, \delta_2 \op{O}) = \frac{1}{2} \oint \df s_1 \oint \df s_2 \dbraket{\delta_1 \op{O}(s_1) \delta_2\op{O}(s_2)}.
\end{equation}
The measure on the boundary is normalized such that $\oint \df s = 1$. We use the same convention as \cite{Shatashvili:1993ps} to distinguish expectation values: $\dbraket{\ \cdot \ }$ is taken with respect to $I$, while $\braket{\ \cdot \ }$ corresponds to $I_0$. In \cite{Witten:1992qy} it was already shown that the form (\ref{symplform}) is closed.

We denote the cohomological vector field at a point $\op{O}$ by $V_\op{O}$. Similar to $b_{-1}$ it is defined through a bulk operator, in this case through the BRST current. Being a tangent vector, $V_\op{O}$ should describe a perturbation from $\op{O}$. In this notation the variation of the BSFT action $S$ reads
\begin{equation}
\delta S = \frac{1}{2} \oint \df s_1 \oint \df s_2 \dbraket{\delta \op{O}(s_1) V_\op{O}(s_2)}.
\end{equation}
We defer the precise realization of $V$ to the next subsection. However, we already stated in the introduction that there are some issues when we take its definition too naively. We need to carefully look for contact terms, since they can generate corrections to $V$. Also, it is not obvious that $V$ squares to zero or that it generates a symmetry of $\omega$ when one includes these contact terms. Nevertheless, it was argued in  \cite{Shatashvili:1993ps} that the latter property holds to first order in the perturbation around a background.

The criterion for $V$ to generate a symmetry of $\omega$ is closedness of $i_{V} \omega$. However, one may be more ambitious by showing that one can integrate $i_{V} \omega$, meaning that we can find an explicit formula for $S$. As explained in the introduction we actually found such an expression for $S$ given by (\ref{Action}). Furthermore, we can infer the nilpotency of $V$ because it generates the minimal model of open string theory which, by construction, forms an $A_\infty$ algebra. The limitation of our derivation is, however, that we restrict ourselves to on-shell conformal perturbations.

On the other hand, since we work in on-shell perturbation theory, we can make two important simplifications. First of all, instead of working on the disc we work on the upper half-plane. All of the following calculations can also be done on the disc, but the formulas of certain conformal transformations are more involved (for example scaling transformations translated to the disc). Conformal symmetry also allows us to fix the positions of the operator insertions in the symplectic structure (\ref{symplform}) to $0$ and $\infty$:
\begin{equation}
\omega(\delta_1\op{O}, \delta_2 \op{O}) = \frac{1}{2} \dbraket{\delta_1 \op{O}(\infty) \delta_2\op{O}(0)},
\end{equation}
where $\op{O}(\infty) = I \circ \op{O}(0)$ with $I(z) = -\frac{1}{z}$.

\subsection{Definition of $b_{-1}$ and the Cohomological Vector Field}

In this subsection we want to find a consistent definition of the operator $b_{-1}$ and the cohomological vector field $V _Q(\mathcal{O})$. In  \cite{Witten:1992qy} the following definitions were proposed
\begin{equation}\label{defQ}
b_{-1}\mathcal{O} = \oint_{C_\alpha} b(v) \mathcal{O} \quad\text{and}\quad V_\mathcal{O} = \oint_{C_\alpha} j_{BRST} \mathcal{O}.
\end{equation}
The contour $C_\alpha$ is the circle of radius $1 - \alpha$, $\alpha \ll 1$ and $v$ is the vector field generating rotations. In this definition the vector field $V$ seems to depend linearly on $\mathcal{O}$.  To find the non-linear dependence, a natural approach is then in which sense these operator equations hold when these objects are accompanied by other operator insertions.

Another way to think of the definition of $V_\mathcal{O}$ is as defining worldsheet BRST transformations of $\op{O}$. Given a background described by $\op{O}$ we get a perturbed background
\begin{equation}
\op{O} + V_{\op{O}} = \op{O} + \{Q,\op{O}\}.
\end{equation}
Again the nonlinear dependence on the background is hidden. The advantage of this viewpoint is, however, that it allows us to think of $V$ as the BRST operator acting on some general boundary operator $A$ in a given background defined by $\mathcal{O}$. In this sense it defines an operator depending linearly on $A$ (but in general nonlinearly on $\op{O}$). The same considerations also work for $b_{-1}$ despite that we do not think of it as defining a vector field.

With the above in mind, given a set of operators $\{A_i\}$, we expect that in 
\begin{equation}\label{BRST1}
\dbraket{\int_{C_{\alpha}}b(v) A_1(x_1)\cdots A_n(x_n)} \ \text{and} \ \dbraket{\int_{C_{\alpha}} j_{BRST} A_1(x_1)\cdots A_n(x_n)},
\end{equation}
we find contributions of the form $b_{-1}A_i$, respectively $\{Q,A_i\}$, for each $A_i$ in the limit $\alpha \rightarrow 0$. For this to make sense the objects $b_{-1}A_i$ and $\{Q,A_i\}$ should have the following three properties:
\begin{enumerate}
\item They should depend linearly on $A_i$.
\item They should be local, meaning that when acting on an operator inserted at some particular point they should produce a new operator at the same point.
\item They should not depend on the other insertions $A_j$ with $ i \ne j$.
\end{enumerate}
Whats coming next is to find expressions for $b_{-1}$ and $\{Q,A_i\}$ satisfying the three properties.

We start with $b_{-1}$. Here the argument is in fact somewhat circular, because this object already enters into the definition of the boundary action. However, we can make a guess and see whether it works. In a conformal background we know the action of $b_{-1}$. It is given by $\oint b(v)A_i$, where the contour runs only around the operator insertion $A_i$. In conformal field theory we know that this just removes a $c$-ghost insertion from the operator $A_i$. Let us  check that this is compatible with (\ref{BRST1}). To a fixed order in the perturbation we have 
\begin{align}
  &\lim_{\alpha \rightarrow 0} \frac{1}{m!} \braket{\int_{C_\alpha} b(v) A_1 \cdots A_n (\int b_{-1}\op{O})^m} \\
=& \sum_{i = 1}^n \frac{1}{m!} \braket{A_1 \cdots b_{-1}A_i \cdots A_n (\int b_{-1}\op{O})^m} + \frac{1}{(m-1)!} \braket{A_1 \cdots A_n \int b_{-1}^2\op{O} (\int b_{-1}\op{O})^{m-1}}, \nonumber
\end{align}
where in this case $b_{-1}$ is taken with respect to the unperturbed background. Since $b^2_{-1} = 0$ the second term in the above expression drops. We find that the boundary interaction does not contribute to the contour integral $\int_{C_\alpha} b(v)$. Therefore,
\begin{equation}
\dbraket{\int_{C_{\alpha}}b(v) A_1(x_1)\cdots A_n(x_n)} =\sum_{i = 1}^n  \dbraket{A_1(x_1) \cdots b_{-1}A_i(x_i) \cdots A_n(x_n)}
\end{equation}
as expected.

\begin{figure}[t]
\centering
\includegraphics[width = .9\textwidth]{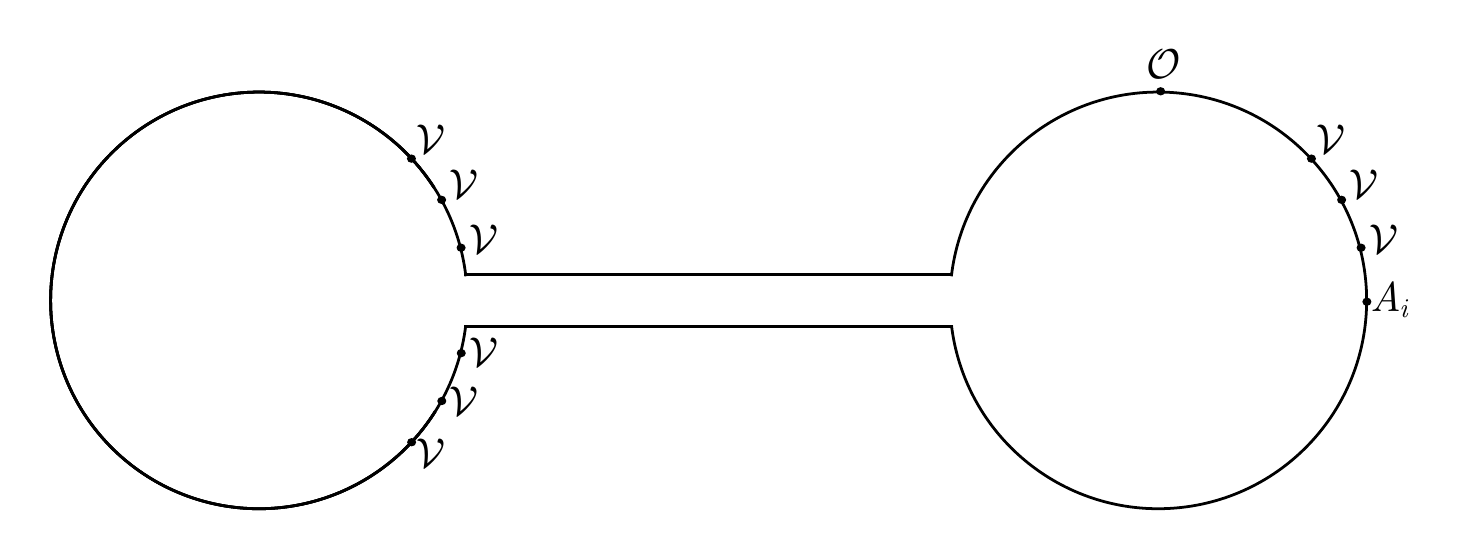}
\caption{A contribution to the boundary of the moduli space giving a correction to the BRST operator. In the region where the boundary operators are close to $A_i$ the disc is equivalent to two discs connected by a narrow neck. Through this, a new local operator is defined inserted in the left disc. Its coefficients with respect to the basis is the expectation value of the right disc. Here $\op{V} = b_{-1}\op{O}$.}
\label{disc_splitting}
\end{figure}

We calculate $V_\op{O}$ along the same lines.  We expand (\ref{BRST1}) to a particular order in the perturbation of the background,
\begin{equation}
\lim_{\alpha \rightarrow 0} \frac{1}{m!} \braket{\int_{C_\alpha} j_{BRST} A_1 \cdots A_n (\int b_{-1}\op{O})^m}.
\end{equation}
Here the BRST charge will now give non-vanishing contributions when acting on both the $A_i$ and $b_{-1}\op{O}$. Since we consider perturbations around a conformal background, for $\op O=0$ we then have  $V=\{Q_0,\cdot\}$, where $Q_0$ is the open string BRST operator. For an on-shell perturbation $\op{O}$ we furthermore have $\{Q_0,b_{-1}\op{O}\} = \partial \op{O}$. Similarly to perturbative open string theory these produce boundary terms in moduli space, parametrized by the positions of the operators $\op{O}$. We distinguish between the following two cases:
\begin{enumerate}
\item Only operators $\op{O}$ are close to each other.
\item Some of the operators $\op{O}$ are close to one of the $A_i$ as in Figure \ref{disc_splitting}.
\end{enumerate}
The first case appears even without operator insertions, thus 
\begin{equation}
0 = \dbraket{\oint_{C_\alpha} j_{BRST}}
\end{equation}
should be absorbed in the definition of $\dbraket{\ \cdot \ }$ and does not contribute to $\{Q,A_i\}$. In contrast the second case arises for each new operator inserted. We will see that they also produce finite contributions described by a new local operator inserted at the position of the operator $A_i$, which can be interpreted as the correction to $\{Q_0,A_i\}$ with respect to the conformal background. So we should find $\{Q, A_i\}= \{Q_0, A_i\} + \{\delta Q, A_i\}$ where $\{\delta Q, A _i\}$ is some local operator that we will determine in the next subsection.  

\subsection{Corrections to the BRST Operator}

\begin{figure}[t]
\centering
\begin{subfigure}[b]{0.29\textwidth}
        \includegraphics[height=4cm]{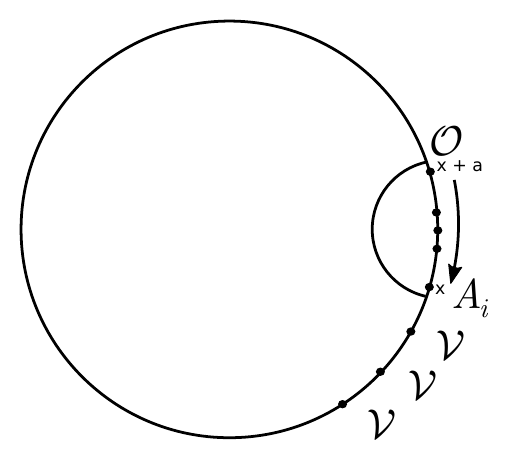}
        \caption{\phantom{.}}
        \label{insertiona}
    \end{subfigure}
\quad
\begin{subfigure}[b]{0.6\textwidth}
        \includegraphics[height=4cm]{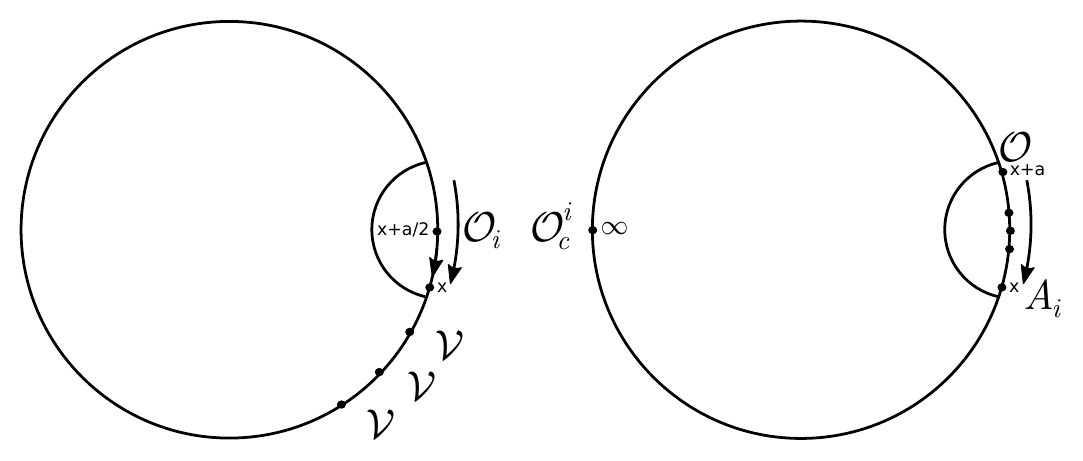}
        \caption{\phantom{.}}
        \label{insertionb}
    \end{subfigure}
 \caption{(a) We split the disc into two along a semicircle of diameter slightly larger than $a$. (b) The same configuration represented by two disjoint discs with insertion of a complete set of operators. When $a \rightarrow 0$, the left disc represents an operator $\op{O}_i$ at $x_i$ in the shifted background, with coefficients computed as an expectation value on the right disc. The arrows represent the limit $a \rightarrow 0$.}
\label{insertion}
\end{figure}

As explained above, the BRST operator receives corrections from contact terms with the boundary interaction. These arise   when the background BRST operator $Q_0$ acts on the boundary interaction:
\begin{equation}\label{Qb}
\{Q_0,e^{\int b_{-1}\op{O}}\} = \int \partial \op{O} e^{\int b_{-1}\op{O}}.
\end{equation}
The integral of the total derivative will give a contribution whenever it collides with another operator. As explained before collisions with operators $b_{-1}\op{O}$ do not appear explicitly. The only relevant ones are collisions with an $A_i$. We focus on one of these collision in (\ref{BRST1}), since by linearity, we just get the sum of every single one. We denote the distance of $\op{O}$ to one of the $A_i$ by $a$, a regulator which eventually will be set to zero, and is chosen so that $\op{O}$ is closer to $A_i$ than to any other $A_j$ for $j \ne i$. Then (\ref{Qb}) contributes to $\{\delta Q,A_i\}$ as 
\begin{equation}\label{abstractQ}
\dbraket{\{\delta Q,A_i\}(x_i) \Omega} = \braket{\op{O}(x_i + a)A_i(x_i) e^{\int b_{-1}\op{O}}\Omega} \pm \braket{A_i(x_i) \op{O}(x_i - a) e^{\int b_{-1}\op{O}}\Omega} ,
\end{equation}
where we denoted all other operator insertions collectively by $\Omega$. The relative sign depends on the statistics of the operator $A_i$. It is not yet clear whether we can bring the right hand side of (\ref{abstractQ}) to the form of the left hand side. So lets move on by focusing on the first summand in (\ref{abstractQ}).  We want to separate the operators $A_i(x_i)$ and $\op{O}(x_i + a)$ from the rest of the operator insertions contained in $\Omega$. We do this by splitting the upper half-plane along a circle of diameter slightly larger than $a$ centered at $x_i +\frac{a}{2}$, so that the two operators are close to the circle. This amounts to inserting a complete set of states (see Figure \ref{insertion})
\begin{equation}\label{idi}
\mathbbm{1} = \sum_j \op{O}_j(x_i +\tfrac{a}{2})\ket{0}\bra{0}\op{O}^j_c(\infty)\,,
\end{equation}
where, for each operator $\op{O}_j$, we define a corresponding conjugate $\op{O}^k_c$ via the condition
\begin{equation}
\braket{\op{O}^k_c(\infty)\op{O}_i(x)} = \delta^k_i.
\end{equation}
By translational invariance the choice of $x$ does not matter.

Of course, we cannot separate the integrated operators whenever they are inserted between $x_i$ and $x_i + a$. We take care of this by splitting each integration range $\mathbb{R} = I_a \cup E_a$, where $I_a = [x_i, x_i + a]$ and $E_a = \mathbb{R}\backslash I_a$. We then find
\begin{gather}\label{dfo}
\braket{\mathcal{O}(x_i + a)A_i(x_i)e^{\int \mathcal{V}}\Omega} = \sum_{k,n \ge 0} \frac{1}{k! n!} \braket{\mathcal{O}(x_i + a)(\int_{I_a} \mathcal{V})^kA_i(x_i)(\int_{E_a} \mathcal{V})^n\Omega} \nonumber \\
= \sum_l \sum_{n \ge 0} \frac{1}{n!}\braket{\mathcal{O}_l(x_i +\tfrac{a}{2})(\int_{E_a} \mathcal{V})^n\Omega} \sum_{k \ge 0} \frac{1}{k!} \braket{\mathcal{O}_c^l(\infty)\mathcal{O}(x_i + a)(\int_{I_a} \mathcal{V})^k A_i(x_i)},
\end{gather}
where we defined $\op{V} = b_{-1}\op{O}$. We can now use a scale transformation by a factor $\frac{1}{a}$ around $x_i$ in the second correlator:
\begin{equation}\label{sa1}
\sum_{k \ge 0} \frac{1}{k!} \braket{\mathcal{O}_c^l(\infty)\mathcal{O}(x_i + a)(\int_{I_a} \mathcal{V})^kA_i(x_i)} = a^{h_l - h_i} \sum_{k \ge 0} \frac{1}{k!} \braket{\mathcal{O}_c^l(\infty)\mathcal{O}(x_i + 1)(\int_{I_1} \mathcal{V})^k A_i(x_i)}.
\end{equation}
We see that for general $h_l$ we get divergences in the limit $a \rightarrow 0$. These divergences can be subtracted by suitable counterterms.  Finite terms appear for $h_l = h_i$. These are universal, as we will see in section \ref{s:offshell}.

Concerning the divergent terms arising in the above limit, recall that also in string theory amplitudes divergences arise when intermediate states have negative dimension, see for example \cite{Witten:2013pra} for a discussion of this. This stems from the fact that the Schwinger parametrization of the propagator
\begin{equation}
\frac{b_0}{L_0} = b_0 \int_0^\infty \df \tau e^{-\tau L_0}
\end{equation}
is only valid for $L_0 > 0$. So we can think of these divergences as artifacts coming from the representation of the amplitude as an integral over positions. 

In the second correlator we have the integral over the region $E_a$, which, for $a \rightarrow 0$, covers the whole integration range. Hence in this limit the correlator becomes
\begin{equation}
\lim_{a \rightarrow 0} \sum_{n \ge 0} \frac{1}{n!}\braket{\mathcal{O}_l(x_i + \tfrac{a}{2})(\int_{E_a} \mathcal{V})^n\Omega} = \dbraket{\mathcal{O}_l(x_i)\Omega}.
\end{equation}
Combining this with the first correlator we find
\begin{equation}\label{sa}
\sum_{\substack{l \\ h_l = h_i}}\dbraket{\mathcal{O}_l(x_i)\Omega} \sum_{k \ge 0} \frac{1}{k!} \braket{\mathcal{O}_c^l(\infty)\mathcal{O}(x_i + 1)(\int_{I_1} \mathcal{V})^k A_i(x_i)}.
\end{equation}
This has the desired form given by the left hand side of (\ref{abstractQ}) upon identifying $\{\delta Q,A_i\}$ with
\begin{equation}
\sum_{\substack{l \\ h_l = h_i}}\mathcal{O}_l(x_i) \sum_{k \ge 0} \frac{1}{k!} \braket{\mathcal{O}_c^l(\infty)\mathcal{O}(x_i + 1)(\int_{I_1} \mathcal{V})^k A_i(x_i)}.
\end{equation}

There is one more contribution coming from $\mathcal{O}$ inserted at $x_i -a$. This basically gives the same answer with different ordering:
\begin{equation}\label{sb}
\sum_{\substack{l \\ h_l = h_i}}\dbraket{\mathcal{O}_l(x_i)\Omega}\sum_{k \ge 0} \frac{1}{k!} \braket{\mathcal{O}_c^l(\infty)A_i(x_i + 1)(\int_{I_1} \mathcal{V})^k \op{O}(x_i)}.
\end{equation}
Comparing (\ref{sa}) and (\ref{sb}) to (\ref{abstractQ}) we find
\begin{equation}\label{sab}
\{\delta Q, A_i\}(x_i) = \sum_{\substack{l \\ h_l = h_i}} \mathcal{O}_l(x_i) \left(\sum_{k \ge 0}\frac{1}{k!}\braket{\mathcal{O}_c^l(\infty)\mathcal{O}(x_i + 1)(\int_{I_1} \mathcal{V})^k A_i(x_i)} \pm [\op{O} \leftrightarrow A_i]\right).
\end{equation}

We now turn to the special case $A_i = \mathcal{O}$ to obtain $V_\op{O}(x)$. We find
\begin{equation}
V_\op{O}(x) = 2 \sum_{\substack{l \\ h_l = 0}} \op{O}_l(x) \left(\sum_{k \ge 0}\frac{1}{k!}\braket{\mathcal{O}_c^l(\infty)\mathcal{O}(x + 1)(\int_{I_1} \mathcal{V})^k \op{O}(x)}\right)
\end{equation}  
which corresponds to (\ref{sab}) with the plus sign, since $\op{O}$ is odd. We now see that the coefficients of the vector field are exactly the on-shell string S-matrices.

\subsection{On-shell Action} \label{onshellaction}

Let us now return to the action
\begin{equation}\label{dsf}
\delta S = \frac{1}{2} \dbraket{\delta \op{O}(\infty) V_\op{O}(0)}.
\end{equation}
Recall again that $\dbraket{\, \cdot \,}$ denotes the expectation value including the perturbed boundary interaction. From the results in the last subsection, on-shell deformations of the boundary worldsheet action lead to
\begin{equation}
\delta S = \lim_{a \rightarrow 0} \frac{1}{2}\dbraket{\delta\op{O}(\infty)\op{O}(a) \op{O}(0)} + \lim_{a \rightarrow 0} \frac{1}{2}\dbraket{\delta\op{O}(\infty)\op{O}(0) \op{O}(-a)}.
\end{equation}
By translational invariance we can rewrite this as
\begin{equation}\label{dS}
\delta S = \lim_{a \rightarrow 0} \dbraket{\delta \op{O}(\infty) \op{O}(a) \op{O}(0)}.
\end{equation}
The expectation value on the r.h.s. of (\ref{dS}) is  actually independent of $a$ by scaling invariance symmetry, which allows us to fix $a = 1$ without loss of generality.
Note that so far we did not assume any concrete renormalization prescription, in analogy to the definition of string amplitudes in perturbative string theory. The action can then be integrated. We find
\begin{equation}
S = \sum_{n \ge 3} \frac{1}{n} \frac{1}{(n - 3)!} \braket{\op{O}(\infty)\op{O}(1)\op{O}(0)\left(\int_{-\infty}^\infty b_{-1}\op{O}\right)^n} =: \frac{1}{2}\sum_{n \ge 3} \frac{1}{n} \mathcal{V}_S^{(n)}(\op{O},...,\op{O})
\end{equation} by noting that each $\mathcal{V}^{n}$ is totally symmetric in the $\op{O}$, a property we derive in the appendix. We see that the action is just a sum of perturbative $S$-matrices. In principle one is free to choose how to deal with divergences (for example using the $i\varepsilon$ prescription as explained in \cite{Witten:2013pra}).

\subsection{A Simple Off-Shell Example}
The key property that allowed us to express the corrections to the BRST operator $Q_0$ induced by the background is identity (\ref{Qb}). For generic off-shell deformations (\ref{Qb}) receives corrections which, in turn, are not given by contact terms. In this case we are not able to determine the corrections $Q_0$ to all orders.\footnote{To first order in the perturbation the correction for generic perturbations was found in \cite{Shatashvili:1993ps}.} An exception to this is the tachyon at zero momentum around a conformal theory with Neumann boundary conditions.  The vertex operator for this background shift is given by $\mathcal{O}(\theta) = T c(\theta)$. This example is somewhat trivial to compute because $b_{-1}\mathcal{O} = T$ and so the background adds no terms beyond linear order to the cohomological vector field:
\begin{equation}
V_T = \{Q_0,\mathcal{O}\} = T c\partial c.
\end{equation}
The differential of the action is then 
\begin{equation}
\df S(T) = \frac{1}{2} \df T Te^{T} \oint \frac{\df \theta_1}{2 \pi} \frac{\df \theta_2}{2 \pi} \braket{c(\theta_1) c\partial c(\theta_2)}.
\end{equation}
Normalizing the ghost 3-point function as 
\begin{equation}
\braket{c(\theta_1) c(\theta_2) c(\theta_3)} = 2(\sin(\theta_1 - \theta_2) + \sin(\theta_2 - \theta_3) + \sin(\theta_3 - \theta_1)).
\end{equation}
we find
\begin{equation}\label{tachyon}
\df S(T) = -\df T T e^{T}.
\end{equation}
For completeness we also give the integrated action
\begin{equation}
S(T) = (1 - T) e^T.
\end{equation}
This is the tachyon action first obtained in \cite{Gerasimov:2000zp}. 
%
%

\subsection{Superstring}
The extension to on-shell states in the NS-sector of the open superstring is straight forward.\footnote{Boundary superstring theory has been previously considered for example in \cite{Kutasov:2000aq}, \cite{Marino:2001qc} and \cite{Niarchos:2001si}.} The background field is naturally integrated over supermoduli space which means that $ \mathcal{V}$ is in the $0$-picture. At the same time it is natural to take $\op{O}(0)$ and $ \op{O}(\infty)$ in the $-1$ picture since they are not integrated in (\ref{dsf}).\footnote{In Witten's original definition $ \op{O}$ and  $\delta \op{O}$ are integrated as well. It is less clear how to extend this definition to supermoduli space.} With this the key identity equation (\ref{Qb}) still holds for the superstring. Furthermore, since $ \op{O}$ and  $\delta \op{O}$ are in the $-1$ picture ($\delta \op{O}$ is contained in $\Omega$ in (\ref{dfo})), this requires that $ \op{O}_j(x_i + \tfrac{a}{2})$ and $\op{O}^j_c(\infty)$ in (\ref{idi}) are in the  $-1$ picture as well. Consequently, (\ref{sa}) and (\ref{sb}) reproduce the superstring $S$-matrix with the correct picture assignments. Recall also that, on-shell the precise location of the picture $-1$ operators is irrelevant as long as the global picture number is correct. 

The extension to the Ramond sector is less clear. Indeed, this requires two Ramond fields in the $-\frac12$ picture and one NS field in the $-1$ picture, or four of the Ramond fields in the $-\frac12$ picture. There does not seem to be a natural way to incorporate this into the definition (\ref{dsf}). Of course, since Ramond fields represent space-time fermions, it is not natural to have non-vanishing Ramond fields in the background. On the other hand, perturbative string amplitudes involving $2n$ fermions are generically non-vanishing. Thus, if BSFT is to realize the minimal model for super string theory the BSFT action should be non-vanishing for an arbitrary even number of Ramond fields. An alternative possibility is that BSFT is minimal only with respect to the NS sector whereas the NS-fields coupling to fermions are not integrated out. This is not in contradiction with what we found before since the NS-sector is a closed subsector of the theory. In that case we would expect vertices with at most two Ramond fields. Then one might postulate that the three unintegrated fields appearing in (\ref{dS}) carry picture $-1$, $-\frac12$, $-\frac12$ respectively. While this might be consistent on-shell, this construction is nevertheless not very natural. 

\section{Alternative Definition of the Cohomological Vector Field}

\label{s:offshell}

For generic off-shell perturbations the construction described above does not generalize directly, because the correction to the cohomological vector field will not be given by boundary terms, since the bulk BRST current does not generate total derivatives when acting on generic off-shell perturbations. In view of this we explore an alternative definition of $V$ through
\begin{equation}
V_\mathcal{O}(0) = \{Q,\mathcal{O}(0)\} = \lim_{a \rightarrow 0} \oint_{\gamma_a(0)} j_{BRST} \mathcal{O}(0),
\end{equation}
where we take $\gamma_a(0)$ to be the semicircle around $0$ with radius $a$, so the difference is merely a choice in the contour. Because the contour approaches the point $0$ in the limit $a \rightarrow 0$, we expect that this limit ensures that $V_\mathcal{O}(0)$ will again be a local operator inserted at $0$. This is in contrast to the original definition, where the contour approached the whole boundary. Also this new definition is closer to $V_\mathcal{O}$ being defined through the worldsheet BRST charge $Q$, since the definition is just that of the latter.

\subsection{On-Shell Perturbations}

We argue on the level of the action. As in section 3, $\oint_{C_a(0)}j_{BRST}$ will act on $\mathcal{O}(0)$ as well as operators coming from the background, but this time only on those which are inside the circle of radius $a$. Therefore we split the integral along the boundary according to $\mathbb{R} = E_a \cup [-a,a]$:
\begin{align}
\delta S =& \frac{1}{2}\dbraket{\delta\mathcal{O}(\infty) \{Q,\mathcal{O}\}} =\frac{1}{2} \sum_{m,n \ge 0} \frac{1}{m! n!} \braket{\delta \mathcal{O} (\infty) (\int_{E_a} b_{-1}\mathcal{O})^{m} \oint_{C_a} j_{BRST} (\int_{-a}^a b_{-1}\mathcal{O})^n \mathcal{O}(0)} \nonumber \\
		 =& \frac{1}{2} \sum_{m,n \ge 0} \frac{1}{m! n!} \braket{\delta \mathcal{O} (\infty) (\int_{E_a} b_{-1}\mathcal{O})^{m}(\int_{-a}^a b_{-1}\mathcal{O})^n \int_{-a}^a \df x \partial \mathcal{O}(x) \mathcal{O}(0)} \nonumber \\
		 =& \frac{1}{2}\int_{-a}^a \df x \, \partial_x \dbraket{\delta \mathcal{O}(\infty) \mathcal{O}(x) \mathcal{O}(0)} \equiv \frac{1}{2}\int_{-a}^a \df x \, \partial_x F(x).
\end{align}
Note that the function $F$ is constant for $\op{O}$ on-shell due to $SL(2,\op{R})$ invariance. However, as we noted before, it picks up a minus sign when we move $\mathcal{O}(x)$ past $\mathcal{O}(0)$. Therefore, we can write $F(x) = \text{sgn}(x)F(1)$. Thus 
\begin{equation}
\delta S = \dbraket{\delta \mathcal{O}(\infty) \mathcal{O}(1)\mathcal{O}(0)},
\end{equation}
so on-shell the new definition reproduces the result of section \ref{onshellaction}.

\subsection{General Off-Shell Perturbations}

For arbitrary off-shell perturbations we could not find an explicit expression for the cohomological vector field $V$. However, we will argue that the coefficients are restricted by the resonance condition given in \cite{Shatashvili:1993ps}. On-shell we found that when we expand the cohomological vector field in the spacetime fields,
\begin{equation}
V^i = \sum_{n \ge 1} V^i_{\;\;i_1 ... i_n} \phi^{i_1} \cdots \phi^{i_n},
\end{equation}
the coefficients are restricted with respect to the scaling dimension of our operator basis $\mathcal{O}_i$. We expect this to hold also in the off-shell case. So
\begin{equation}\label{Offres}
V^i_{\;\;i_1 ... i_n} \propto a^{h_i - \sum_{k = 1}^n h_{i_k}}.
\end{equation}

To discuss which of the coefficients $V^i_{i_1...i_n}$ are universal, the use of the Poincar\'{e}-Dulac theorem proves to be powerful, as was pointed out in \citep{Shatashvili:1993ps}. For convenience we restate its consequences. Suppose we have a vector field expanded in some coordinates $x^i$ around a critical point, $V(x^i = 0) = 0$. This means that we can write
\begin{equation}
V^i(x) = A^i_j x^j + \sum_{n \ge 2} V^i_{i_1...i_n}x^{i_1}\cdots x^{i_n}.
\end{equation} 
For simplicity we assume that the linear term is diagonal, $A^i_j = h_i \delta^i_j$, which will be sufficient for our discussion. The theorem can also be stated for nondiagonal $A^i_j$. The assertion of the theorem is then that we can find a coordinate transformation
\begin{equation}
y^i(x) = x^i + \text{higher orders}
\end{equation}
so that all higher order terms $V^i_{i_1...i_n}$ satisfying
\begin{equation}
h_i \ne \sum_{k = 2}^n h_{i_k}
\end{equation}
vanish in the $y$ coordinate.

In our case the entries of the diagonal matrix $A^i_j$ are the conformal weights of the fields. We see from (\ref{Offres}) the divergent terms are those where
\begin{equation}
h_i < \sum_{k = 2}^n h_{i_k},
\end{equation}
and therefore can be removed by field redefinitions according to the Poincar\'{e}-Dulac theorem. The cutoff-independent terms, however, are exactly those satisfying the ``resonance condition''
\begin{equation}\label{res2}
h_i = \sum_{k = 2}^n h_{i_k}.
\end{equation}
\begin{figure}[t]
\centering
 \includegraphics[width= .7\textwidth]{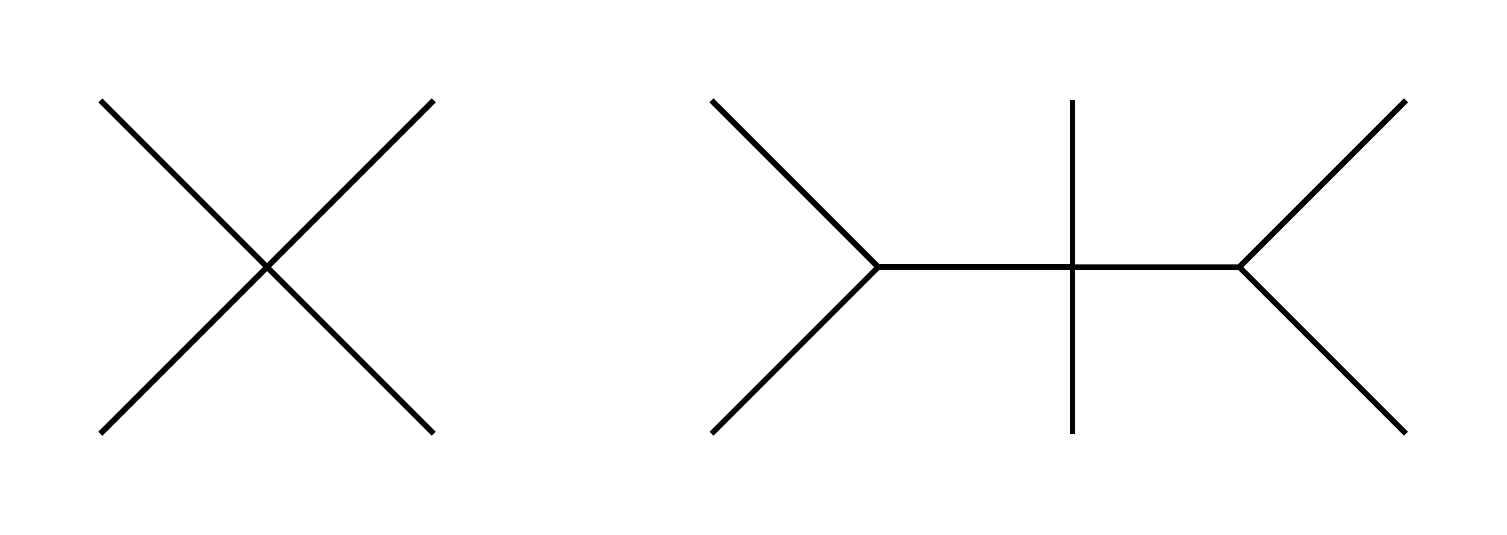}
 \caption{The left diagram consists only of a vertex. It has only on-shell states attached, so it is generated by BSFT. The right diagram has one vertex with two internal lines and two vertices with only a single internal line. Because BSFT does not contain the latter type of vertex, BSFT does not generate this diagram.}
\label{tree}
\end{figure}

The resonance condition has an important consequence for the cohomological vector field
\begin{equation}
V^i=h_i \phi^i + \sum_{n \ge 1} V^i_{\;\;i_1 ... i_n} \phi^{i_1}\cdots \phi^{i_n}.
\end{equation} 
If $\phi^i$ is off-shell ($h_i \ne 0$), then at least one of the $\phi^{i_k}$ is also off-shell due to (\ref{res2}). Thus an off-shell field cannot be sourced by on-shell fields only. A consequence of this is that there are no vertices in BSFT that couple on-shell states to a single off-shell state. Therefore all vertices have to contain either zero, two or more internal lines. However, tree-level Feynman diagrams, apart from those which are the vertices themselves, contain at least two vertices with only a single propagator attached (see Figure \ref{tree}).  Thus, the minimal model of BSFT is again that of OSFT. Moreover, we see that classical OSFT is equivalent to BSFT on the perturbative level, because they predict the same scattering amplitudes.

\section{Conclusions}
We were able to show that for on-shell deformations of the background, the expansion of the cohomological vector field $V$, of BSFT, produces the S-matrices of perturbative open string theory which, in turn, is the minimal model of open string field theory. Thus, BSFT, when restricted to on-shell perturbations, realizes the minimal model of OSFT. This clarifies the relation between OSFT and BSFT: On-shell, BSFT generates the minimal model of OSFT. This also explains why BSFT has no propagator, a fact that was often considered an odd property of BSFT. In addition this equivalence also implies that $V$ is nilpotent in some neighborhood around a background, which is a non-trivial result. The extension of this equivalence can easily be extended to the NS sector of the open superstring. 

The inclusion of the Ramond sector is, however, problematic, since there is no natural way to assign picture to the background in this sector. It would be interesting to find a consistent generalization of (\ref{dsf}) to supermoduli space. Another limitation of our approach is that it does not easily allow to give a detailed description of the cohomological vector field off-shell. In particular, we can not guarantee that $V$ squares to zero off-shell, and thus, whether BSFT does indeed define a consistent off-shell BV action in some neighborhood of a conformal background. Progress in this direction would certainly be helpful in order to decide whether BSFT can resolve some of the infrared issues in string perturbation theory.

\section*{Acknowledgments}
We would like to thank Sebastian Konopka for helpful discussions. Part of this work was presented at the String Field Theory Conference 2018 at the HIT in India. CC wants to thank the organizers for the chance to present this work and also for the hospitality provided during the conference. This work  was supported by the DFG Transregional Collaborative Research Centre TRR 33 and the DFG cluster of excellence "Origin and Structure of the Universe".

\appendix
\section{Cyclic Symmetry}
In this section we want to explain the well known but important cyclic symmetry property of open string amplitudes.

We begin with the first non-trivial case, the four-point scattering amplitude: Suppose we are given four string states represented by conformally invariant operators $\op{O}_i$, where $i \in \{1,...,4\}$. Those can have arbitrary ghost number and hence arbitrary statistics, but we assume that they are multiplied by a string field $\phi^i$ so that the total degree is equal to one, just as in the case of ordinary open string field theory. We assume that the punctures, where the operators are inserted, are in a given order on the boundary of the disc. The moduli space of this punctured disc is then one dimensional.

The usual representation of the four-point S-matrix with given cyclic order $(1234)$ is
\begin{equation}
\mathcal{V}^{(4)}(1,2,3,4) = \braket{\op{O}_1(\infty) \op{O}_2(1) (\int_0^1 b_{-1}\op{O}_3) \op{O}_4(0)}
\end{equation}
with $b_{-1} = b(v)$ and $v$ the vector field generating translations along the boundary. The choice of integrated operator in the above expression is arbitrary. We could as well have chosen to integrate $\op{O}_4$. Therefore
\begin{equation}
\mathcal{V}^{(4)}(1,2,3,4) = \braket{\op{O}_1(\infty) \op{O}_2(1) \op{O}_3 (0)(\int_{-\infty}^0 b_{-1}\op{O}_4)}.
\end{equation}
Using $\text{SL}(2,\mathbb{R})$ invariance we can write this in the standard representation
\begin{equation}
\mathcal{V}^{(4)}(1,2,3,4) = \braket{\op{O}_2(\infty) \op{O}_3(1) (\int_0^1 b_{-1}\op{O}_4) \op{O}_1(0)} = \mathcal{V}^{(4)}(2,3,4,1).
\end{equation}
The overall sign of this transformation is $+1$, because $\op{O}_1$ has to commute with one even and two odd objects. This shows that the four-point S-matrix has indeed cyclic symmetry.

We also introduce the symmetric correlator $\mathcal{V}_S^{(4)}$, which is just the sum over all cyclically inequivalent correlators (meaning that they cannot be mapped to each other by a cyclic permutation). There is a nice and compact way to represent this symmetric correlator. Notice that we have $|S^4/C^4| = |S^3| = 6$ inequivalent cyclic orderings. Half of them can be written as
\begin{equation}
\braket{\op{O}_1(\infty) \op{O}_2(1) \op{O}_3(0) (\int_{-\infty}^{\infty} b_{-1}\op{O}_4)}.
\end{equation}
The integration range splits into three parts: $(-\infty,\infty) = (-\infty,0] \cup [0,1] \cup [1,\infty)$. This produces $(1,2,3,4)$, $(1,2,4,3)$ and $(1,4,3,2)$. We can get the other half by adding the same amplitude with index $2$ and $3$ swapped. Therefore
\begin{equation}
\mathcal{V}^{(4)}_S (\op{O}_1,\op{O}_2,\op{O}_3,\op{O}_4) = \braket{\op{O}_1(\infty) \op{O}_2(1) \op{O}_3(0) (\int_{-\infty}^{\infty} b_{-1}\op{O}_4)} + (2 \leftrightarrow 3).
\end{equation}

Higher order correlators can be generated in the same way:
\begin{align}
\mathcal{V}^{(n)}_S (\op{O}_1,...,\op{O}_n) =& \frac{1}{(n-3)!} \braket{\op{O}_1(\infty) \op{O}_2(1) \op{O}_3(0) \int_{-\infty}^{\infty}b_{-1}\op{O}_4 \cdots \int_{-\infty}^{\infty}b_{-1}\op{O}_n} \nonumber \\
+& (2 \leftrightarrow 3).
\end{align}
We can again split the integration ranges according to the ordering of the operators (with reference point at infinity). It is then clear that the first half produces all orderings of the tuple $(1,2,...,n-1,n)$ with $1$ fixed and $2$ always to the left of $3$. Hence we can again generate all cyclically inequivalent orderings by adding the amplitude with just $2$ and $3$ swapped. There is also no overcounting because we fix the position of $1$ to be at infinity.

\cleardoublepage
\addcontentsline{toc}{section}{References}
\bibliography{lit}
\bibliographystyle{JHEP}


\end{document}